\documentclass[final,1p,11pt]{elsarticle}

\usepackage{epsfig}
\usepackage{array,tabularx,epsfig,mathrsfs,graphicx,rotating}
\usepackage{ifthen}
\usepackage{amsfonts}
\usepackage{ragged2e}
\PassOptionsToPackage{hyphens}{url}
\usepackage[hyphens]{url}
\usepackage{hyperref}
\usepackage{listings}
\usepackage{subfigure}
\usepackage{epstopdf}
\usepackage{color}
\usepackage{float}
\usepackage{verbatim}
\usepackage{color,soul}
\usepackage{subfigure}

\usepackage[normalem]{ulem} % either use this (simple) or
\usepackage{soul} % use this (many fancier options)
\usepackage{amsmath,amssymb}

\let\originallesssim\lesssim
\let\originalgtrsim\gtrsim

\DeclareRobustCommand{\lesssim}{%
  \mathrel{\mathpalette\lowersim\originallesssim}%
}
\DeclareRobustCommand{\gtrsim}{%
  \mathrel{\mathpalette\lowersim\originalgtrsim}%
}

\makeatletter
\newcommand{\lowersim}[2]{%
  \sbox\z@{$#1<$}%
  \raisebox{-\dimexpr\height-\ht\z@}{$\m@th#1#2$}%
}
\makeatother

\hypersetup{
  colorlinks=true,
  linkcolor=blue,
  citecolor=blue,
  urlcolor=blue
}

\graphicspath{{figs/}}

\newcommand{\beq}{\begin{equation}}
\newcommand{\eeq}{\end{equation}}

\chardef\til=126

\journal{}

\begin{document}
\hfill ANL-HEP-159872

\definecolor{mygreen}{rgb}{0,0.6,0} \definecolor{mygray}{rgb}{0.5,0.5,0.5} \definecolor{mymauve}{rgb}{0.58,0,0.82}

\lstset{ %
 backgroundcolor=\color{white},   % choose the background color; you must add \usepackage{color} or \usepackage{xcolor}
 basicstyle=\footnotesize,        % the size of the fonts that are used for the code
 breakatwhitespace=false,         % sets if automatic breaks should only happen at whitespace
 breaklines=true,                 % sets automatic line breaking
 captionpos=b,                    % sets the caption-position to bottom
 commentstyle=\color{mygreen},    % comment style
 deletekeywords={...},            % if you want to delete keywords from the given language
 escapeinside={\%*}{*)},          % if you want to add LaTeX within your code
 extendedchars=true,              % lets you use non-ASCII characters; for 8-bits encodings only, does not work with UTF-8
 keepspaces=true,                 % keeps spaces in text, useful for keeping indentation of code (possibly needs columns=flexible)
 frame=tb,
 keywordstyle=\color{blue},       % keyword style
 language=Python,                 % the language of the code
 otherkeywords={*,...},            % if you want to add more keywords to the set
 rulecolor=\color{black},         % if not set, the frame-color may be changed on line-breaks within not-black text (e.g. comments (green here))
 showspaces=false,                % show spaces everywhere adding particular underscores; it overrides 'showstringspaces'
 showstringspaces=false,          % underline spaces within strings only
 showtabs=false,                  % show tabs within strings adding particular underscores
 stepnumber=2,                    % the step between two line-numbers. If it's 1, each line will be numbered
 stringstyle=\color{mymauve},     % string literal style
 tabsize=2,                        % sets default tabsize to 2 spaces
 title=\lstname,                   % show the filename of files included with \lstinputlisting; also try caption instead of title
 numberstyle=\footnotesize,
 basicstyle=\small,
 basewidth={0.5em,0.5em}
}

% AVK linenumbers

\begin{frontmatter}

\title{
Physics potential of timing layers in future collider detectors}
%%%%%%%%%%%%%%%%%%%%%%%%%%%%%%%%%%%%%%%%%%%%%%%%%%%%%%%%%%%%%%%

\author[add1]{S.V.~Chekanov}
\ead{chekanov@anl.gov}

\author[addDuke]{A.V.~Kotwal}
\ead{ashutosh.kotwal@duke.edu}

\author[add3]{C.-H. Yeh}
\ead{a9510130375@gmail.com}

\author[add3]{S.-S.~Yu}
\ead{syu@cern.ch}

\address[add1]{
HEP Division, Argonne National Laboratory,
9700 S.~Cass Avenue,
Argonne, IL 60439, USA.
}

\address[add3]{
Department of Physics and Center for High Energy and High Field Physics, 
National Central University, Chung-Li, Taoyuan City 32001, Taiwan
}

\address[addDuke]{
Department of Physics, Duke University, USA
}

\begin{abstract}
The physics potential of timing layers with a few tens of pico-second resolution in the 
calorimeters of future collider detectors is explored.
These studies show how such layers can be used for particle identification and  illustrate the potential for detecting  
new event signatures originating from physics beyond the standard model. 
\end{abstract}

\end{frontmatter}

%%%%%%%%%%%%%%%%%%%%%%%%%%%%%%%%%%%%%%%%%%%%%%%%%%%%%%%%%%%%%%%%%%
\section{Introduction}

Future particle colliders such as CLIC \cite{Linssen:1425915}, the International Linear Collider (ILC) \cite{Behnke:2013xla}, the high-energy LHC (HE-LHC),
and $pp$ colliders of the European initiative, FCC-hh~\cite{Benedikt:2206376} and the Chinese initiative (CEPC \cite{CEPCStudyGroup:2018ghi} and  SppC~\cite{Tang:2015qga}) 
will motivate  high-precision measurements of particles and jets 
at large transverse momenta. 
Timing information in these experiments  can be used to improve particle and jet reconstruction and to suppress backgrounds.
For example, high-precision timing will be beneficial for $b$-tagging for  post-LHC experiments. 
At CLIC and FCC, high-precision time stamping of calorimeter energy deposits will be essential for
background rejection (i.e. fake energy deposits) and pile-up mitigation.
Precise timing information improves reconstruction of particle-flow objects by reducing overlap 
of out-of-time energy showers in highly-granular calorimeters.

Timing layers can be used for detection of long-lived particles and identification of standard model (SM) particles. 
Conceptual design reports for future experiments have not yet fully explored 
the benefits of the time-of-flight (TOF) measurements with calorimeters having tens-of-picosecond resolution.

In this paper we investigate the benefits of timing layers with resolution in the range 10~ps -- 1~ns. 
The resolution of 1~ns is standard for existing and planned calorimeters~\cite{Linssen:1425915,Behnke:2013xla,Benedikt:2206376,Tang:2015qga}, and is used as a benchmark for comparisons with the more challenging 10 -- 20~ps  resolution devices. 
In addition, we investigate the capabilities of timing layers for identification of heavy stable particles which may be produced due to 
 physics beyond the 
standard model (BSM). These studies can help shape the requirements for future calorimeters, which were already 
outlined in the CPAD report~\cite{Ahmed:2019sim} that emphasized the need to develop fast calorimetric readouts.
  
%%%%%%%%%%%%%%%%%%%%%%%%%%%%%%
\section{Proposal}
%%%%%%%%%%%%%%%%%%%%%%%%%%%%%

A generic design of calorimeters for future collider experiments 
is based on two main characteristics: (1) high-granularity  electromagnetic (ECAL) and hadronic (HCAL) 
calorimeters with cell sizes ranging from $3\times 3$~mm$^2$ to $5\times 5$~cm$^2$.
(2) timing with a nanosecond precision that improves background rejection, vertex association, and detection of new particles. 
According to the CPAD report~\cite{Ahmed:2019sim}, the development of ``picosecond time resolution'' for calorimeters is one 
 of the critical needs. 
Presently,  high-granularity calorimeters with $>1$ million channels and with tens of picosecond resolution represent a 
significant challenge due to the large cost.

As part of the HL-LHC upgrade program, CMS and ATLAS experiments are designing high-precision timing detectors with 
 resolution of about 30~ps~\cite{Collaboration:2623663,CMS:2667167}. They are based on silicon sensors that add an 
 extra ``time dimension’’ to event reconstruction. 
In the case of the CMS High Granularity Calorimeter \cite{Collaboration:2293646},
six  million channels of the endcap calorimeter require a dedicated front-end with $\simeq$ 20 picosecond time binning.

As mentioned above, timing capabilities have not been fully explored for detectors beyond the HL-LHC upgrade, such as 
 the ECAL and HCAL of the future detectors for the CLIC, CEPC, FCC and SppC experiments. 
It is considered an expensive option for the many millions of channels of these highly granular detectors,
assuming that high-precision timing is implemented for all calorimeter cells.  
This opens an opportunity to investigate a cost-effective ``timing layer'', with time resolution better than 30~ps,  
to be installed in front of barrel and endcap calorimeters.

In this paper we  will investigate the physics advantages of such timing layers in the post-LHC experiments. 
Typically, thin detectors in front of calorimeters are called ``preshower'', and were used in the  ZEUS, CDF, ATLAS and CMS experiments.
Their design goal is to count the number of charged
particles in order to correct for upstream energy losses. The timing information of minimum-ionizing particles (MIP) has 
not been used for particle identifications or event timing. 
Unlike the standard preshower detectors, we propose to not only count MIP signals, but also measure 
 high-precision timing and position of the MIPs.
This timing detector will have a similar granularity as the proposed high-granularity ECALs, 
but will have sensor and readout technology that are best suited for time-stamping of MIP signals.
Our proposal is to enclose the ECALs with two timing layers, one before the first EM layer, and the other after the 
last ECAL layer (but before the first HCAL layer). The two timing layers allow a robust calculation of time stamps by correlating 
 the position and timing of the particles passing through the ECAL.

\begin{figure}
\begin{center}
   \includegraphics[width=0.8\textwidth]{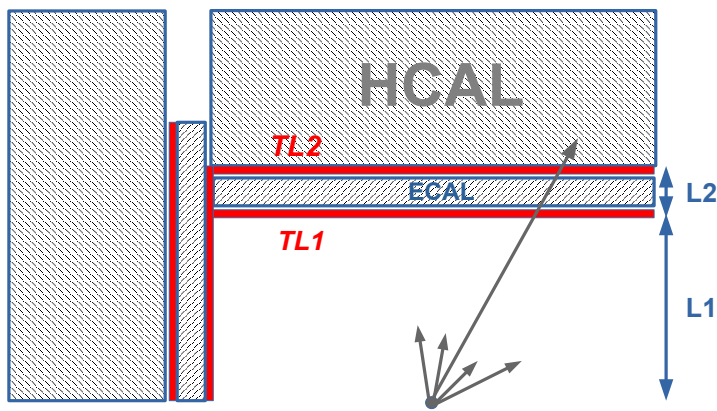}\hfill
\end{center}
\caption{Example positions of thin timing layers for a generic detector. The timing detectors  enclose the 
electromagnetic calorimeter, allowing  a reliable calculation of the  MIP signals with a timing resolution of the order of 10~ps.}
\label{fig:eff_rad}
\end{figure}

We explore this idea using a semi-analytical approach and using Monte Carlo simulations. 
A schematic representation of the positions of the timing layers for a generic detector geometry is shown in Fig.~\ref{fig:eff_rad}. 
In the following, the first timing layer (closest to the interaction point) will be called TL1, while the second timing layer after the
 ECAL will be called TL2.

There are several reasons why the second timing layer (TL2) can be useful:

\begin{itemize}

\item
It can be used to measure the TOF between TL2 and TL1 for identification
of stable massive particles without a known production vertex. This is especially important
for BSM models predicting stable heavy particles
decaying close to the surface of the ECAL, or
in the situations when tracks cannot be used to establish the production vertex for neutral heavy particles.
For proton colliders,  the second layer can mitigate situations when the primary vertex position  
is smeared due to large pileup.

For a typical ECAL based on silicon technology, the distance between TL2 and TL1 is  20 -- 40~cm, depending on the calorimeter design.
It is not immediately obvious that such a small distance can be used for physics measurements. 
A particle traveling at the speed of light can cross this distance within $\sim 1$~ns.
As we will discuss later, this distance is sufficient to provide a large acceptance 
for heavy particle identification  assuming  10 -- 20~ps resolution detectors. 

\item
It allows to correlate the hits with the first layer, and thus provide directionality of the hits. This feature can be useful to
match the hits with the calorimeter cells and to deal with back-scatter hits which are typically arriving from the HCAL at a later time. 

\item
It provides redundancy for the calculation of TOF using the distance from the production vertex determined using tracks.

\end{itemize}

The second layer of the timing detector can be justified if the recorded time difference between the first and the last ECAL  
layers of the electromagnetic showers is not significantly different from that expected from a particle traveling with speed of light.
If the travel time is significantly affected by large fluctuations caused by electromagnetic showers,    
the second timing layer cannot be used effectively.

In order to verify this point, we used a full {\sc geant}4 (version 10.3)~\cite{Allison2016186} simulation 
of the SiFCC detector~\cite{Chekanov:2016ppq} that allows the use of the ECAL hit information.
This detector design has an ECAL built from highly segmented silicon-tungsten cells with transverse size $2 \times 2$~cm$^2$.
The ECAL has 30 layers of tungsten pads with silicon readout,
corresponding to 35~X$_{0}$. The first 20 layers use tungsten of 3~mm thickness, while the last ten layers use tungsten layers of
twice the thickness, and thus have half the  sampling fraction.  
The distance between the centers of the first and last ECAL layers is about 24~cm.  

To check that the time differences between the EM hits in the last and the first ECAL layer is close to the time
required for a particle that travels with the speed of light,
two samples of single pions ($\pi^\pm$) with momenta of 1 and 10~GeV respectively were created. The
pseudorapidity ($\eta$) for all pions was fixed at $\eta=0$ (central region). 
The particles were reconstructed in the ECAL, and the time difference $\Delta T= T_{\mathrm{last}}-T_{\mathrm{first}}$ of the hits 
 between the last and first ECAL layers was calculated.
Only the hits arriving first in time were considered since the electronics typically register\footnote{The Monte Carlo studies in
 this paper do not include a simulation of calorimeter electronics.} the fastest hits, while slower hits can be saved in pipeline buffers.

\begin{figure}
\begin{center}
   \subfigure[] { 
   \includegraphics[width=0.45\textwidth]{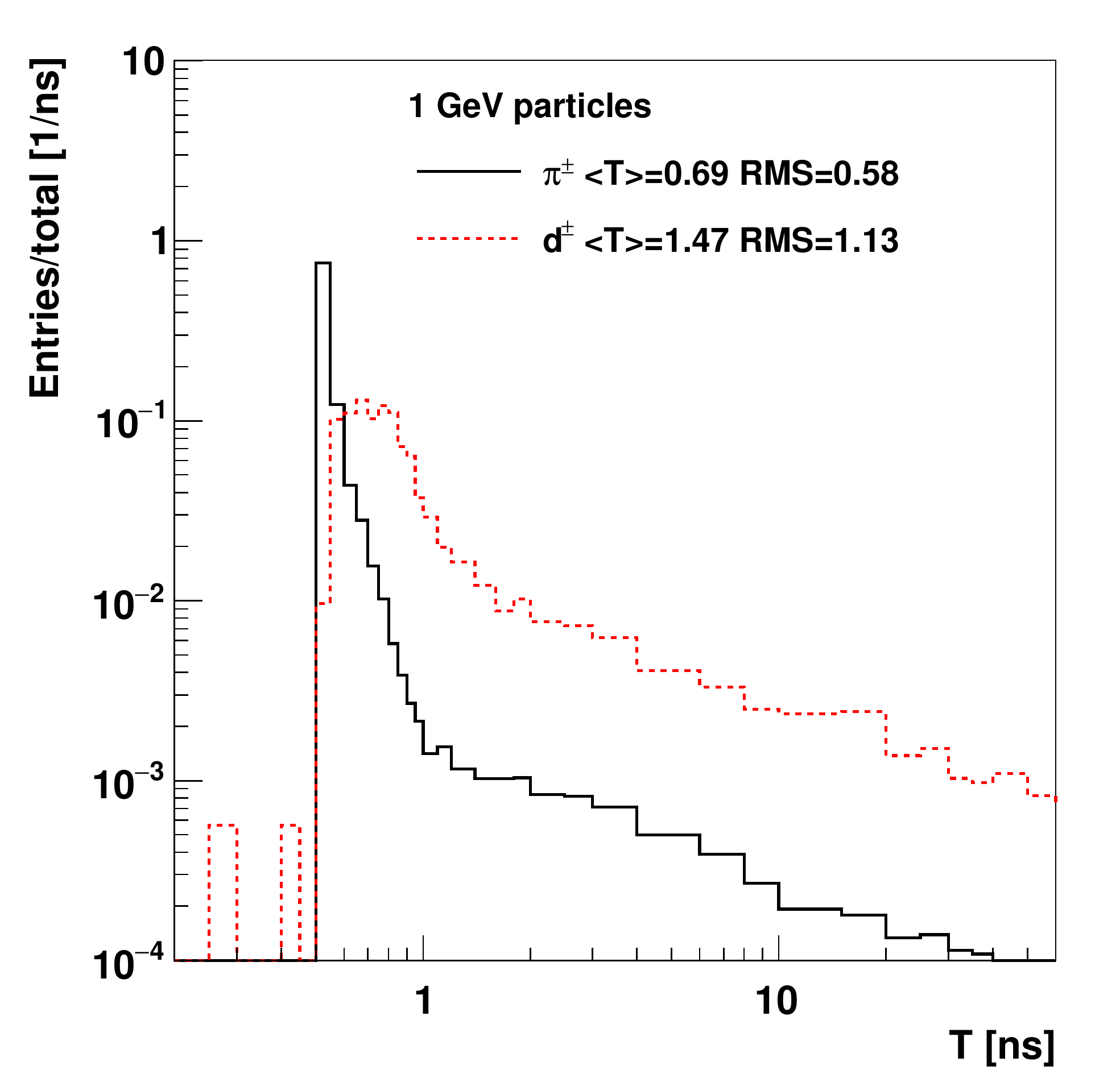}\hfill
   }
   \subfigure[] { 
   \includegraphics[width=0.45\textwidth]{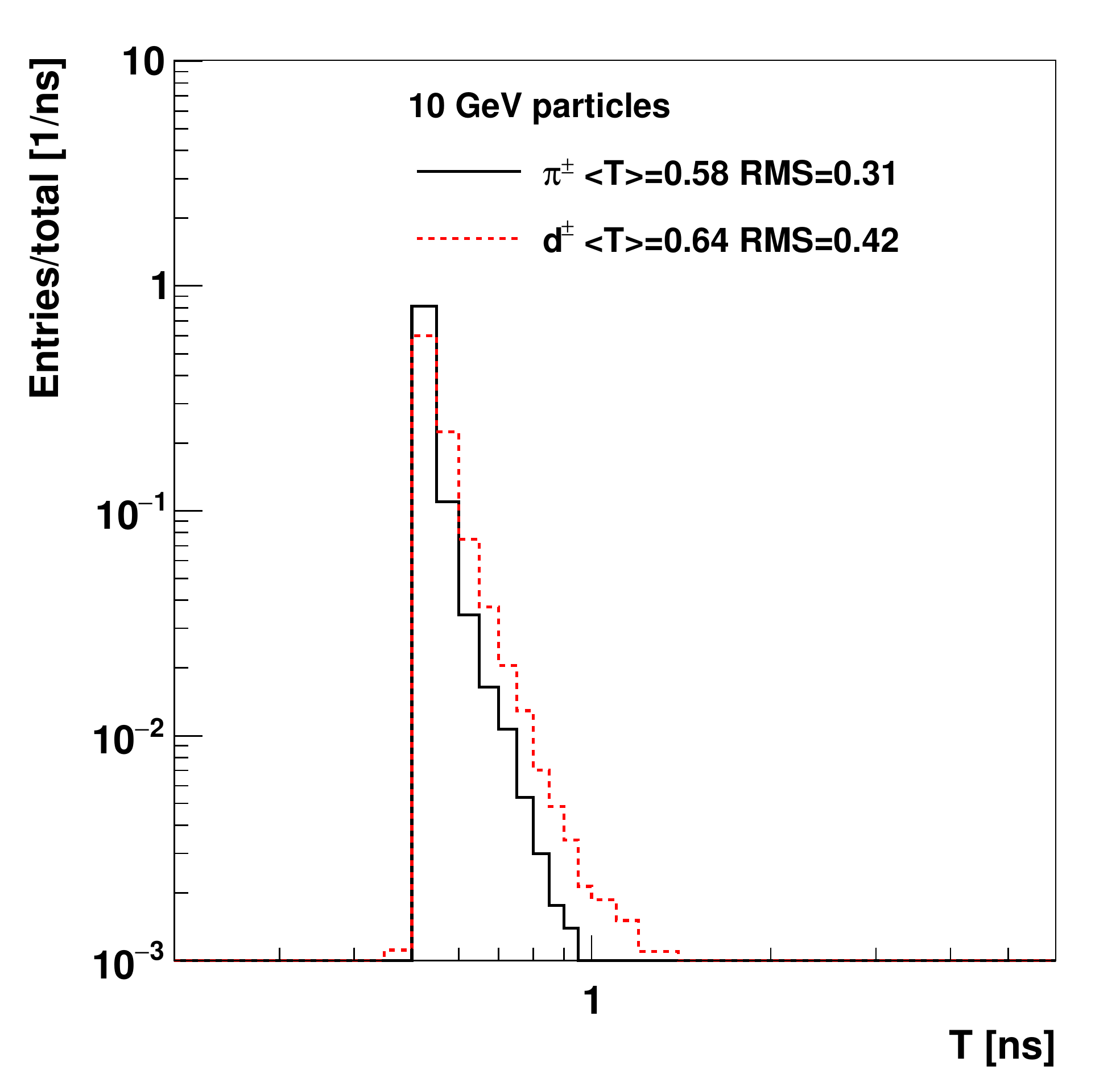}\hfill
   }
\end{center}
\caption{Difference between hit times at the last and first layer of ECAL for single pions and deuterons  with 
 momentum of (a) 1~GeV  and (b) 10 GeV. Only the earliest hits at each layer were considered in the  calculation of the TOF. }
\label{fig:timediff}
\end{figure}

Figure~\ref{fig:timediff} shows the time distribution of the earliest hits 
for 1 and 10~GeV pions. It can be seen that the peak positions of the distributions are smaller 
than 1~ns, as expected for the distance of 24~cm between the last and first ECAL layers.
Therefore, the hits registered by TL1 and TL2 will be considered simultaneous for the
standard 1~ns resolution readout. They will be fully correlated in time, and will be identified as a single crossing particle.

If the resolution of the timing layer is of the order of 10 -- 20~ps, TOF can be used to identify particles
which are heavier than pions. 
To check this, the arrival time of the earliest hits was calculated for (anti)deuterons (denoted as $d^{\pm}$) using the same
simulation setup as before.
Deuterons are 14 times heavier than pions, can be produced in the primary interactions, 
and their interaction with the detector material is well understood.  
If the difference between hit arrival times for  $\pi^\pm$ and $d^{\pm}$ are larger than 20~ps, this   
would indicate a sensitivity of tens-of-picosecond detectors to particles with different masses.

Fig.~\ref{fig:timediff} shows the time distribution of the earliest hits for $d^{\pm}$. 
The distributions are significantly different from the $\pi^{\pm}$ case. According to the simulation, the 1~GeV
(anti)deuterons should be measured with the time delay of 0.7 -- 1.4~ns between the last and first layers.
The value of 0.7~ns was estimated from the peak of the Landau distribution used to fit the $d^{\pm}$ 
curve presented in Fig.~\ref{fig:timediff}(a),
while 1.4~ns was obtained from the mean of this distribution. Even for the most conservative 0.7~ns value, this indicates 
that 1~GeV deuterons can be separated from pions that have 0.5~ns time difference. Such a separation can be observed 
using a tens-of-picosecond detector since the hits produced by the electromagnetic shower
are sufficiently fast.

For particle identification in realistic collider environments, the 1~GeV range momentum range is rather low. 
For the 10~GeV particles shown in Fig.~\ref{fig:timediff}(b), 
no separation between $d^{\pm}$ and $\pi^{\pm}$ can be observed.
This result is totally expected since the mass difference between pions and deuterons is  too small
for effective separation of these particles at a momentum above 10~GeV. 

In summary, we have illustrated that a typical time difference between TL2 and TL1 (which is approximated by the difference
between the last and first ECAL layer) is sufficient for identification of low-mass particles below the GeV-scale in momentum.
As an example, a $d^{\pm}$ can be separated from pions for a momentum less than 1~GeV.
It is expected that particles that are heavier than deuterons can be identified using TOF for a momentum larger than 1~GeV.
In the following, we will abstract from the {\sc geant}4 simulations and calculate the kinematic regions  where the 
 identification of heavy stable  particles using timing layers is possible.

\clearpage 
%%%%%%%%%%%%%%%%%%%%%%%%%%%%%%%%%%%%%%%%%%%%%%%%%%%%%%%%%%%%%%%%%%

%%%%%%%%%%%%%% sections 
\section{Timing layers for single particles}

Now we discuss the kinematic regions  relevant for  TOF measurements of SM and BSM particles. Instead of the full {\sc geant}4 simulations, we will
use a semi-analytic approach.  
 
To estimate the separation power between different mass hypotheses, we calculate the mass and momentum for which one can achieve a separation 
 significance higher than $3\sigma$ (or p-value$<0.3$\%). 
If there are two particles with mass $m$ and a reference (fixed) mass $m_F$, respectively, the $3\sigma$ separation can be 
achieved for this condition~\cite{Cerri:2018rkm}:
\begin{equation}
\frac{L}{c \sigma_{\textsc{TOF}}}\left|\sqrt{1+\frac{m^2}{p^2}} - \sqrt{1+\frac{m_F^2}{p^2}}\right| > 3
\label{eqTOF}
\end{equation}
where $p$ is the momentum of a particle with mass $m$, $L$  is the length of the particle's trajectory, 
and $\sigma_{TOF}$ is the
resolution  of the timing layer that measures the TOF.

\begin{figure}
\begin{center}
   \subfigure[Neutrons] {
   \includegraphics[width=0.45\textwidth]{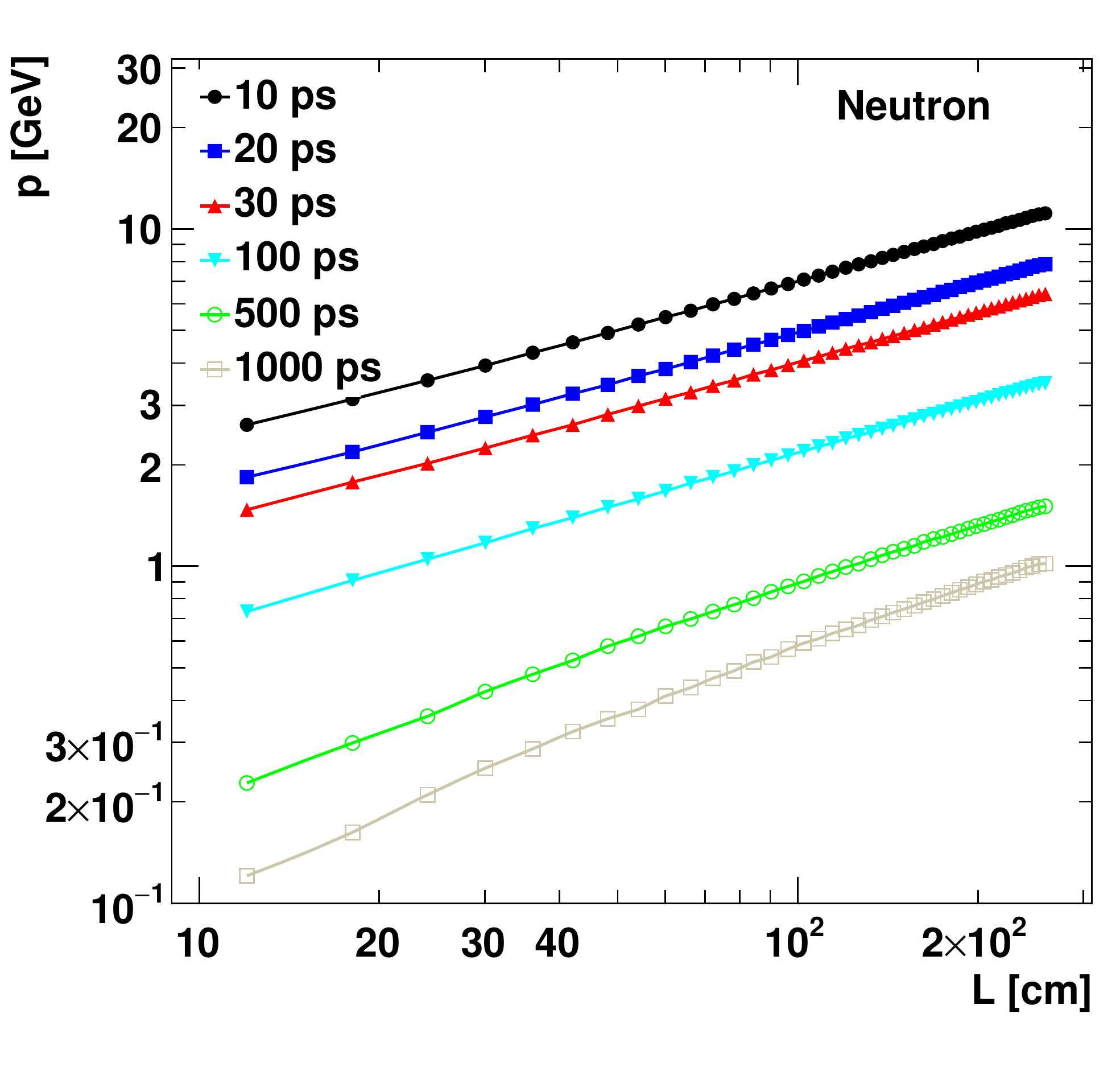}
   }
      \subfigure[$K$-mesons] {
   \includegraphics[width=0.45\textwidth]{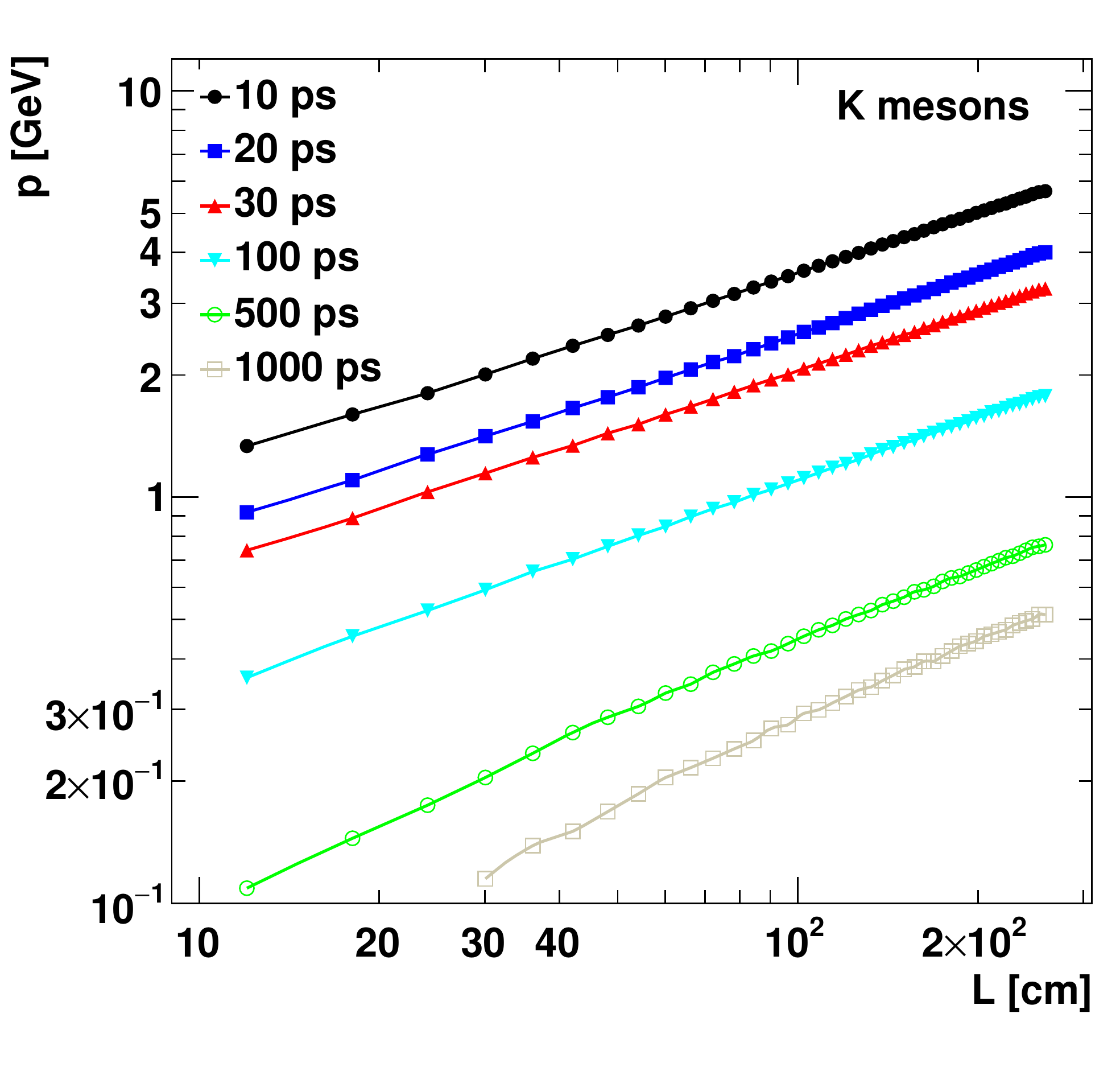}\hfill
   }
\end{center}
\caption{
The $3\sigma$ separation from the pion-mass hypothesis for (a) neutrons and (b) kaons as a function of the length of the particle's trajectory $L$ 
and the momentum $p$. The lines show extrapolated results between the calculations indicated by the symbols.  
}
\label{fig:singleparticles}
\end{figure}

Figure~\ref{fig:singleparticles} shows the $3\sigma$ separation from the pion
mass hypothesis ($m_F=m_{\pi}$) using the procedure discussed  in~\cite{Cerri:2018rkm}. The 
calculations are performed for several values of resolution of the timing layer, ranging from 10~ps to 1~ns,
as a function of $L$ and $p$. For a 20~ps detector and a typical travel 
distance $L\sim 1.5-2$~m from the production vertex to the ECAL, neutrons and protons can be separated from the pion hypothesis up to $p \approx 7$~GeV. 
The separation of kaons from pions can be performed up to 3~GeV.
This momentum range should be sufficient for a reliable particle 
identification in a momentum range adequate for some physics studies focused on
single-particle reconstruction (such as B-meson physics).
This can also be used for jets that are dominated
by particles in this momentum range.
For a detector  with 1~ns resolution, the separation can only be possible  up to  300 -- 500~MeV. This is smaller than the 
minimum particle momentum of $\approx$0.5~GeV considered for high-energy proton colliders.
Therefore, a timing layer with 1~ns resolution cannot be used for particle identification in such experiments.

\begin{figure}
\begin{center}
   \subfigure[for $L=2$~m] {
   \includegraphics[width=0.45\textwidth]{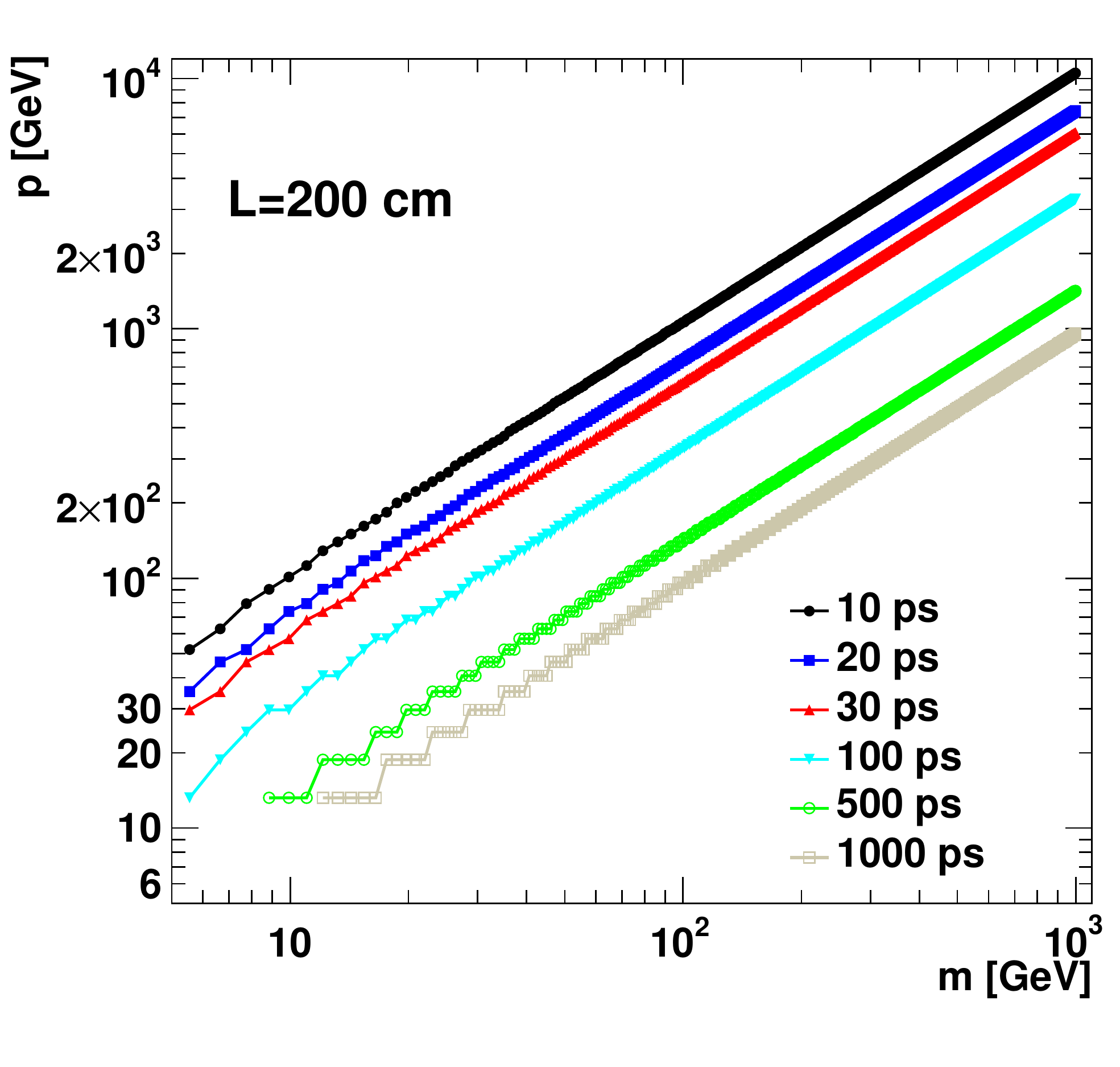}
   }
   \subfigure[for $L=0.2$~m] {
   \includegraphics[width=0.45\textwidth]{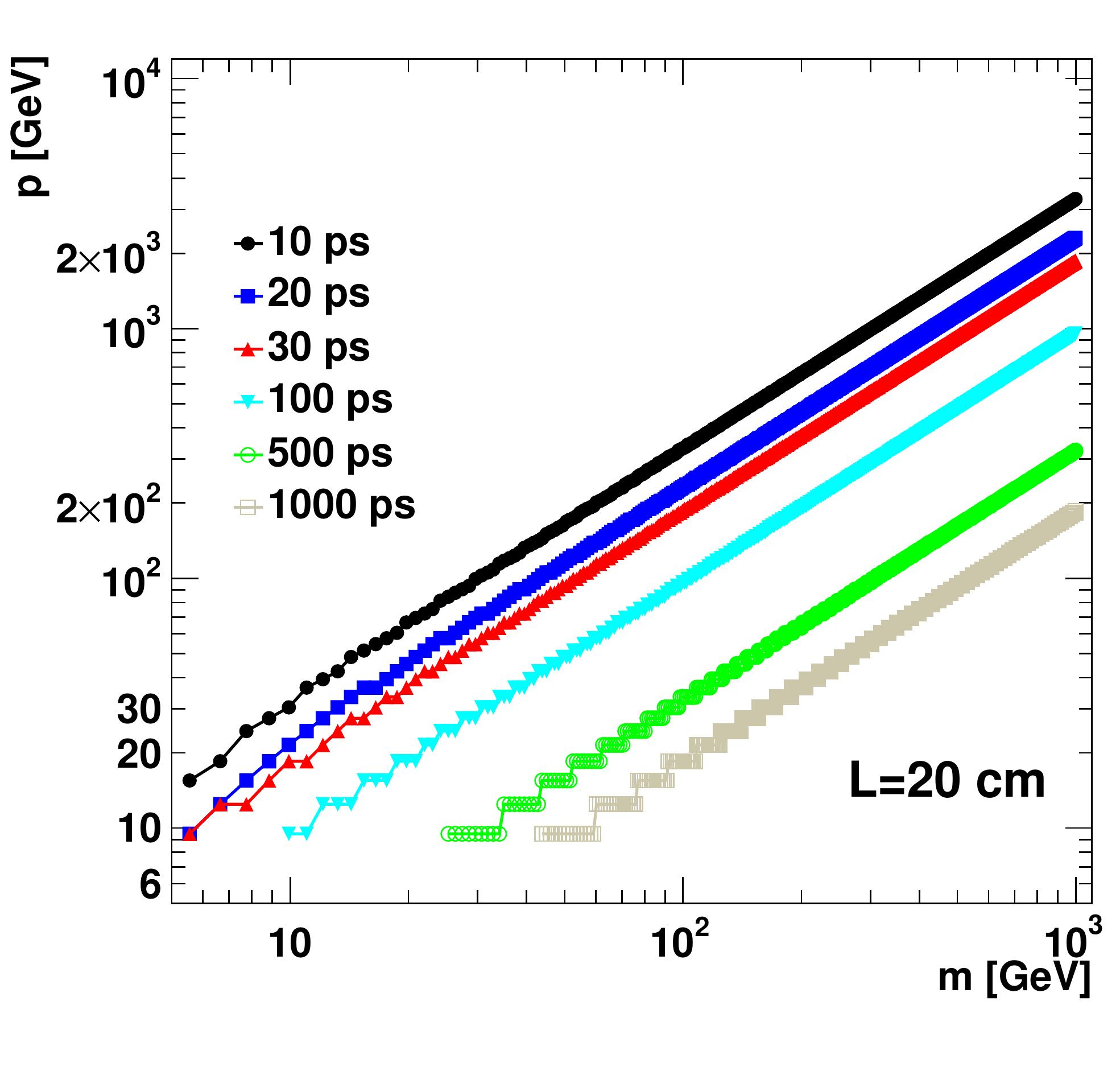}\hfill
   }
\end{center}
\caption{
The $3\sigma$ separation between heavy particles and $\alpha$ particles assuming timing layers with different resolutions for TOF, and using (a) $L=2$~m and (b) $L=0.2$~m.
The first value of $L$ is the typical distance
from the interaction vertex to the first layer TL1, while the second value is the typical  distance
between the two timing layers enclosing an ECAL based on the silicon technology.
}
\label{fig:signgleBSM}
\end{figure}

Having discussed the rather classical cases of discriminating  neutrons, protons and kaons from the pion hypothesis,
let us turn to the BSM searches for heavy particles.
The largest SM  backgrounds for light BSM  particles are primary protons and neutrons.
Other stable particles that can be produced by secondary interactions in the 
detector material or the beam pipe are deuterons and $\alpha$ particles. 
Although the $\alpha$ particle rate is  low since they stop easily in detector material,
it may still represent background for rare BSM particle searches.  
Therefore, we choose  $m_F=m_{\alpha}\simeq 3.73$~GeV  as reference\footnote{We clarify that the choice of $\alpha$ particles as the reference mass is
arbitrary and is only motivated by our attempt to check the $3\sigma$ separation in the momentum range $p<10$~GeV.} in  Eq.~\ref{eqTOF}, and evaluate the
$3\sigma$ separation for a wide range of masses and momenta above 4~GeV.
For many planned experiments the distance between the $pp$ collision point and the first layer of the ECAL is 
$1.5-2.5$~m. Therefore we use $L=2$~m and consider 0.2~m as the separation between the TL2 and TL1 timing layers.

Figure~\ref{fig:signgleBSM} shows the discrimination power for different choices of the timing layer resolution
and the distance $L$ (see Fig.~\ref{fig:eff_rad}).
For $L=2$~m, a stable heavy particle of mass 100~GeV can be discriminated for momentum up to 
700~GeV assuming a 20-ps timing layer,
but only up to 50~GeV using the standard 1~ns resolution.

When  TOF is measured between the layers TL1 and TL2, and assuming a spatial match of the hits, the knowledge of the interaction vertex is not required.
This type of measurement can be beneficial for neutral particles in events with large pile-up (multiple $pp$ collisions).
The identification power when $L=0.2$~m, i.e. the distance between TL2 and TL1, is shown in Fig.~\ref{fig:signgleBSM}(b).
For a stable particle with mass 100~GeV, the identification is possible up to about 200~GeV in momentum. The standard calorimeter with
1~ns resolution can only perform the identification up to $20$~GeV.

\section{Showcase for the Dark QCD model}
\label{darksec}

The arguments discussed above can be illustrated using concrete BSM physics scenarios.
We will consider the ``dark'' QCD model~\cite{Bai:2013xga,Schwaller:2015gea}, which predicts 
the existence of ``emerging'' jets 
that are created in the decays of new long-lived neutral 
particles (dark hadrons), produced in a parton-shower process by dark QCD.
The process includes  two mediators of mass $M_X$ which 
decay promptly to a SM quark and a dark quark. 
The final-state signature consists of four jets with high transverse momenta, with two  
 emerging jets originating from the dark quarks.  

Searches for emerging jets have been performed in $pp$ collisions~\cite{Sirunyan:2018njd} 
by the CMS Collaboration. Such jets contain many displaced
 vertices and multiple tracks with large impact parameters, arising from the decays of the dark pions produced in the dark parton shower.
 Assuming that the mass of the dark pion is 5~GeV,  the signal acceptance using this approach does not exceed 40\% at large masses of the mediators
(see Fig.~4 of~\cite{Sirunyan:2018njd}).
The decay length of the dark pion defines the distance from the $pp$ interaction vertex 
to the point where the jet emerges. 

Alternatively, emerging jets can be reconstructed using calorimeters with high-resolution timing. This method is expected
to have advantages over the track-based method 
for the measurement of dark pions with a large decay length, i.e. in the situations where the tracker has a low
efficiency and resolution since only a few outer layers can be used for track reconstruction.
It was also pointed out~\cite{Schwaller:2015gea} that the emerging jets may have a significant fraction of neutral particles and the reconstruction
using charged tracks can have a low acceptance.

To estimate the performance of the timing layers in reconstructing emerging jets,
we use the same Monte Carlo generator settings as for Ref.~\cite{Sirunyan:2018njd}. 
The $pp$ collision event samples  were  generated with the ``hidden valley'' model framework as 
implemented in  {\sc pythia} 8.2 \cite{Sjostrand:2007gs}
assuming a centre-of-mass energy of 13~TeV and a dark pion mass of 5~GeV. 
The samples were created for different values of the decay distance $c\tau$ of the dark pions.  
The  mass $M_X$ of the mediator was also varied. 

To calculate the detector acceptance, the semi-analytical formalism based on Eq.~\ref{eqTOF} is used. In this relation $L=c\tau$ is the distance traveled
by the dark pion with mass $m$ before it decays to the emerging jet.  
We assume that such emerging jets travel to the surface of the timing layer with speed of light for all values of $m$.
This is expected since the emerging jets consist of light stable SM particles (mostly photons and pions).
  
For the timing layers, the signature of emerging jets is a time delay compared to the other SM jets. The  production vertex
cannot be observed by the timing layers if such jets emerge before TL1.
After events are generated, the weighted averages of the decay distances of all particles that originate from
the dark pions, using the particle momentum as the weight, were  calculated. This decay distance is used
to approximate the decay length, without applying a jet reconstruction algorithm.  
The  calculation for the $3\sigma$ separation assumes $m_F=m_{\alpha}\simeq 3.73$~GeV although this choice can be arbitrary.
This value of $m_F$ is used to give a conservative\footnote{One can argue that the SM jets mainly consist of light-flavour hadrons and photons, 
therefore, $m_F$ should be significantly lower.} estimate of the arrival time of the emerging SM jets.

\begin{figure}
\begin{center}
   \subfigure[10 ps] {
   \includegraphics[width=0.39\textwidth]{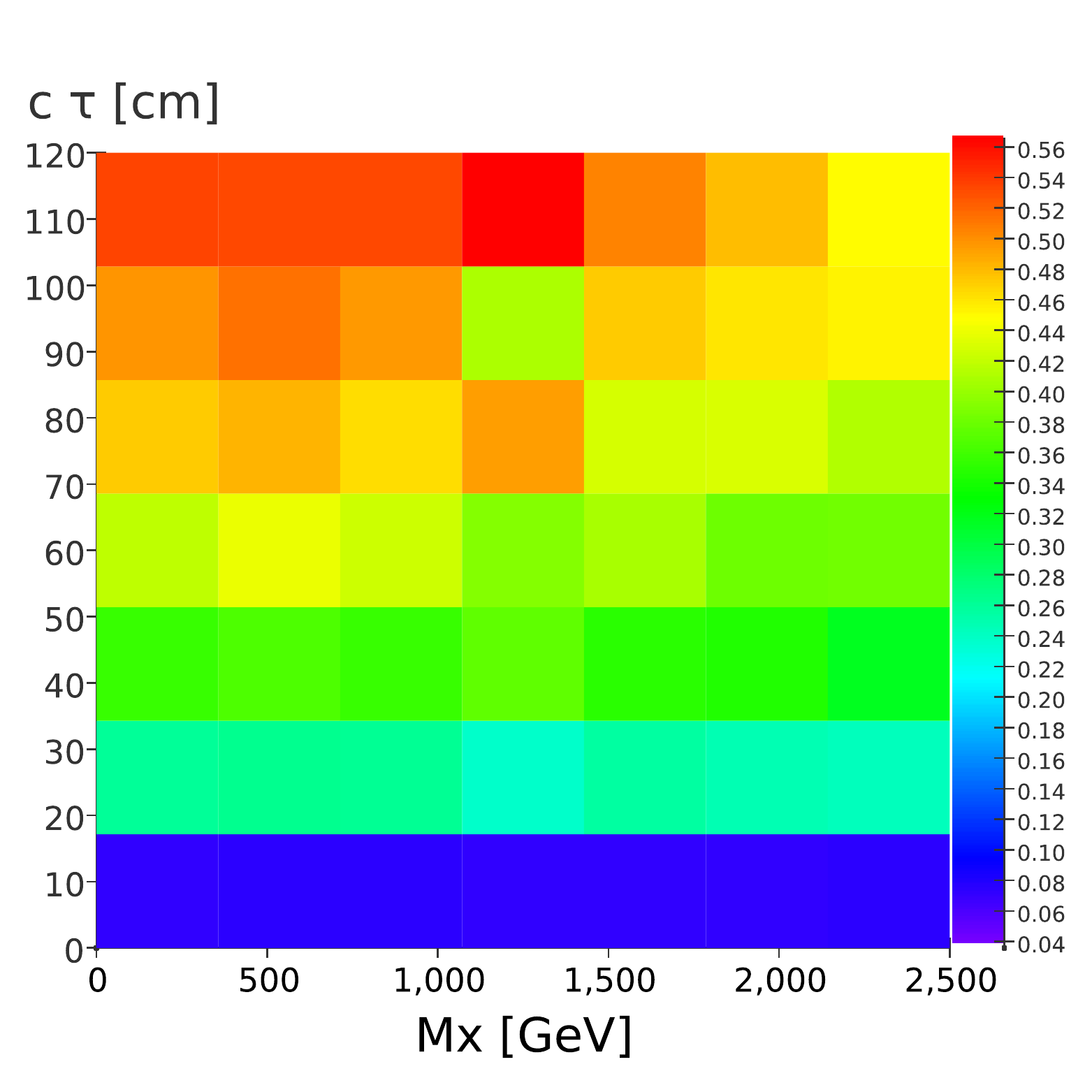}
   }
   \subfigure[20 ps] {
   \includegraphics[width=0.39\textwidth]{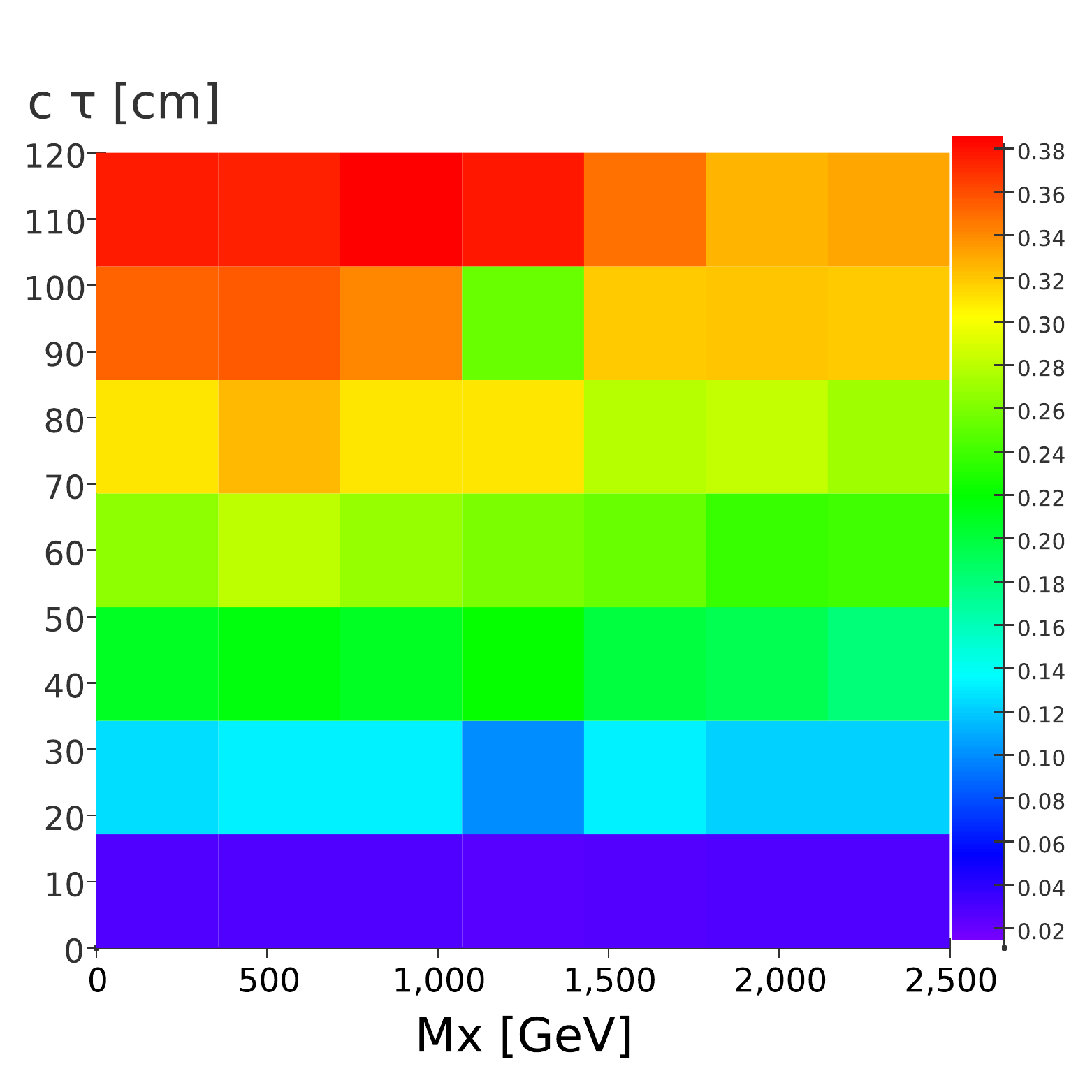}\hfill
   }

   \subfigure[30 ps] {
   \includegraphics[width=0.39\textwidth]{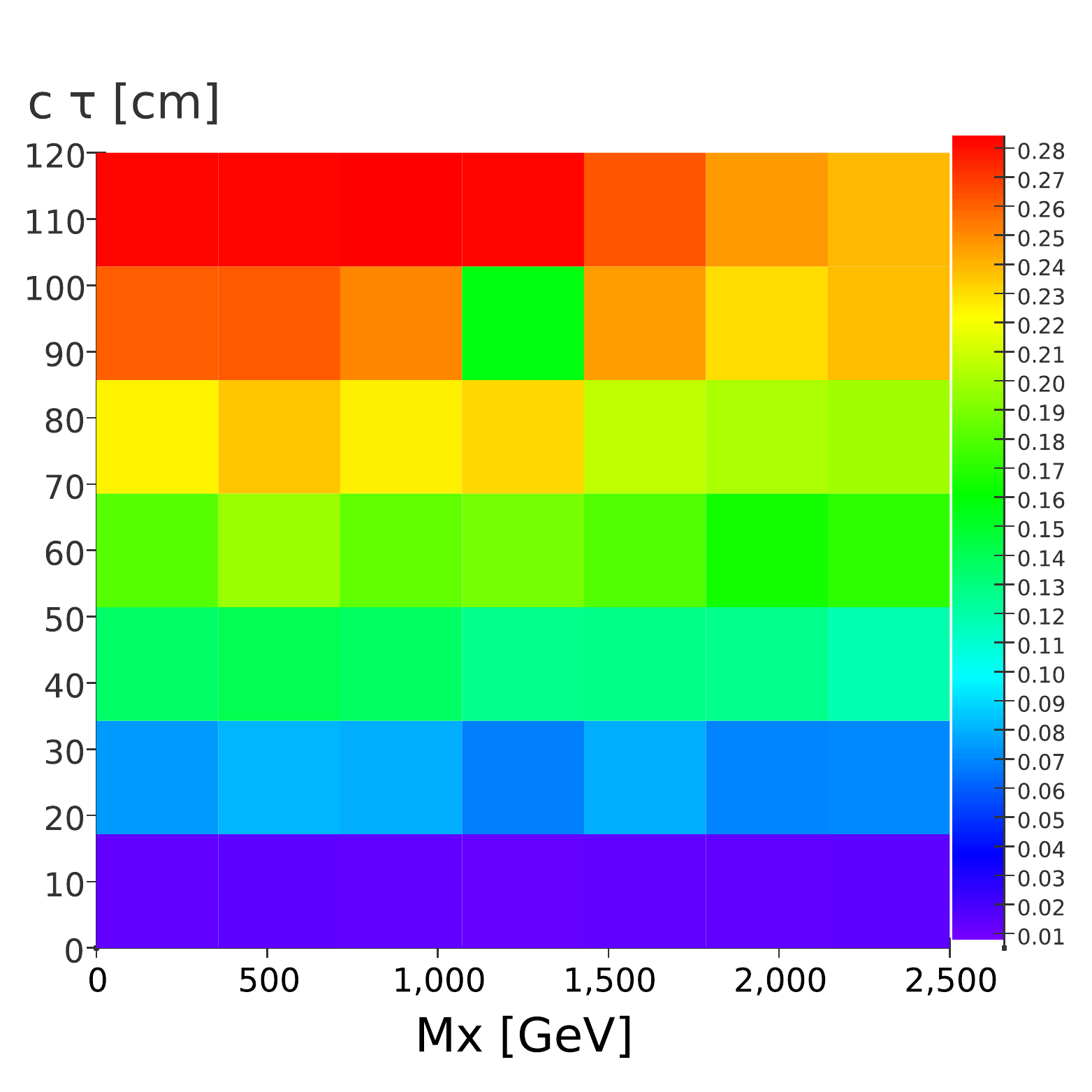}
   }
   \subfigure[100 ps] {
   \includegraphics[width=0.39\textwidth]{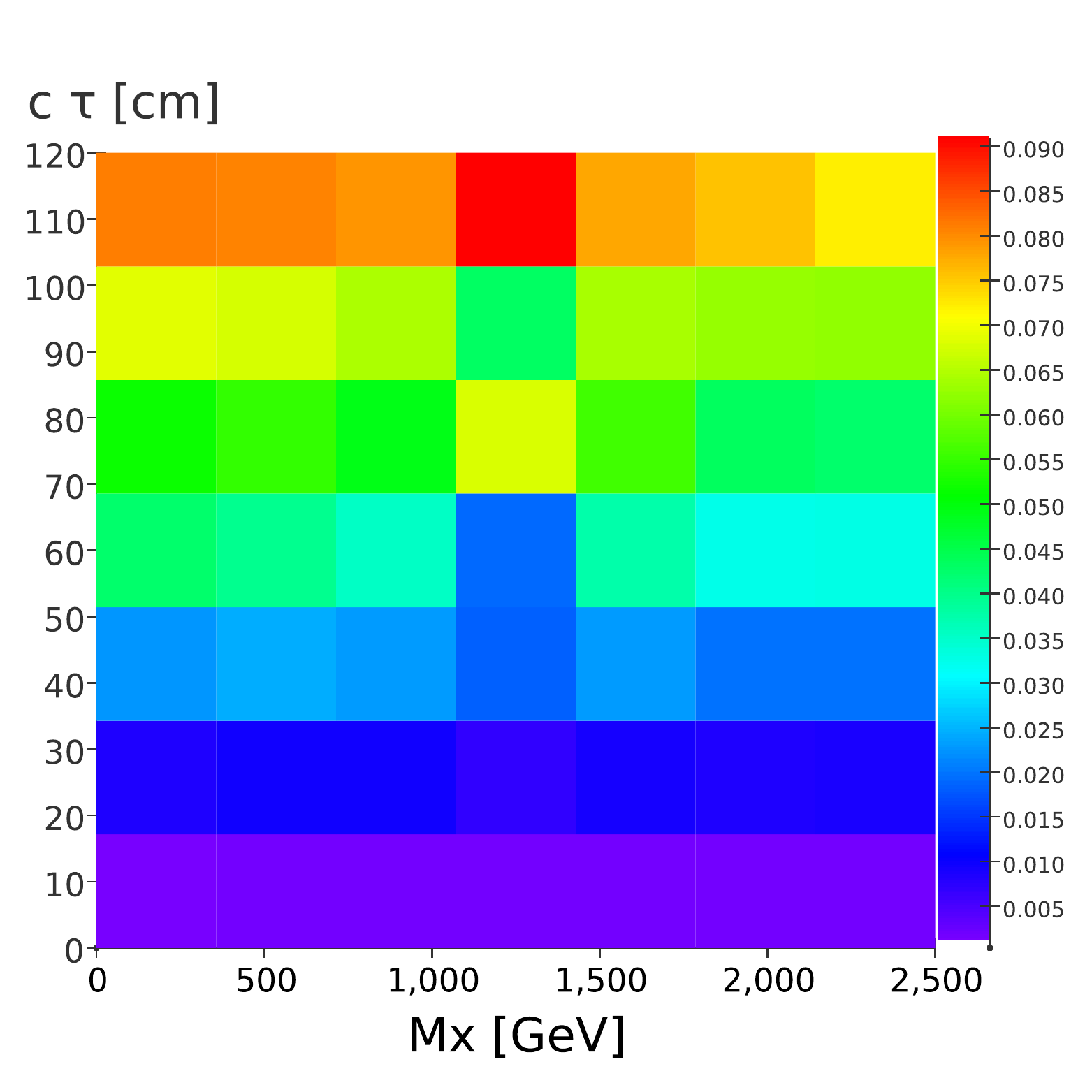}\hfill
   }

   \subfigure[500 ps] {
   \includegraphics[width=0.39\textwidth]{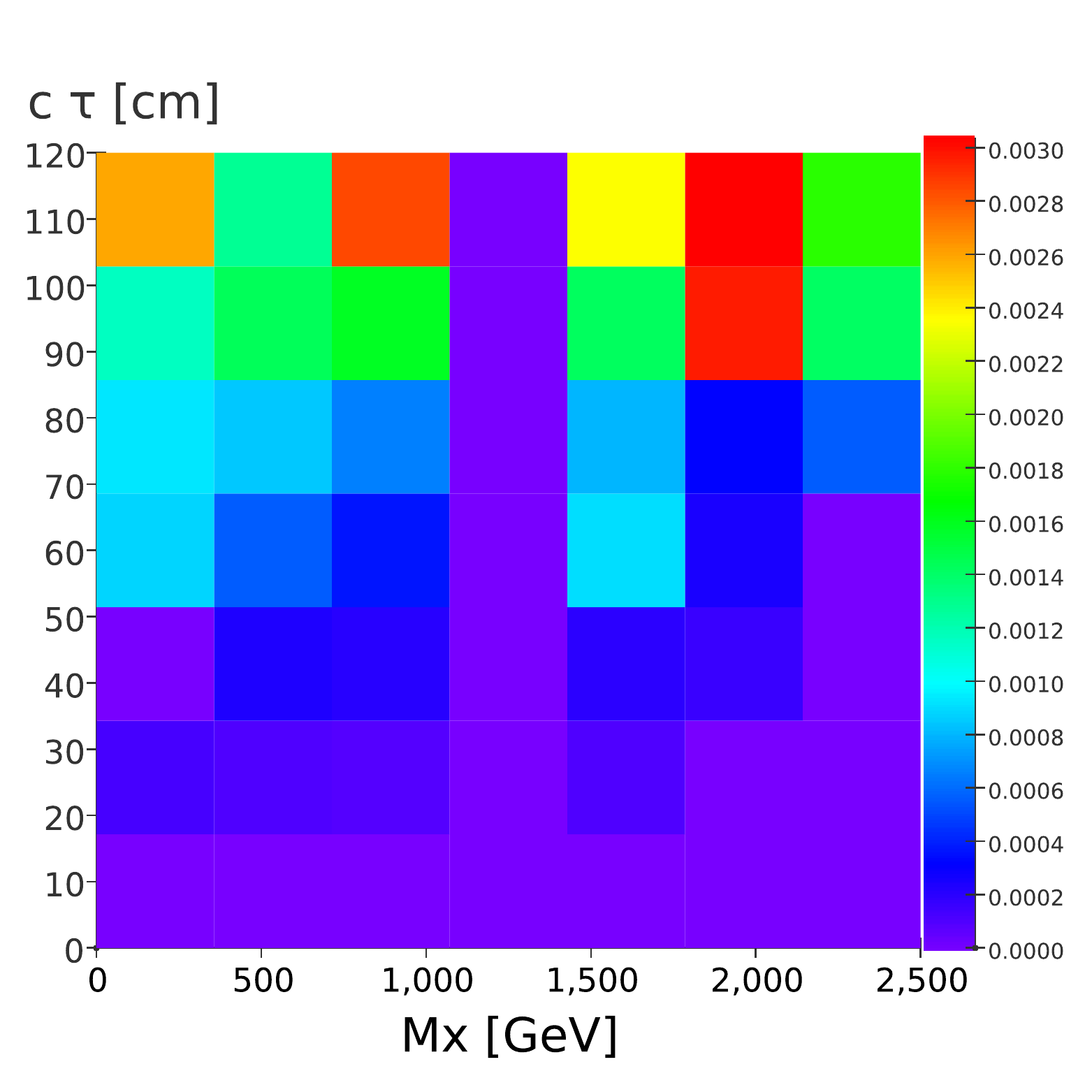}
   }
   \subfigure[1000 ps] {
   \includegraphics[width=0.39\textwidth]{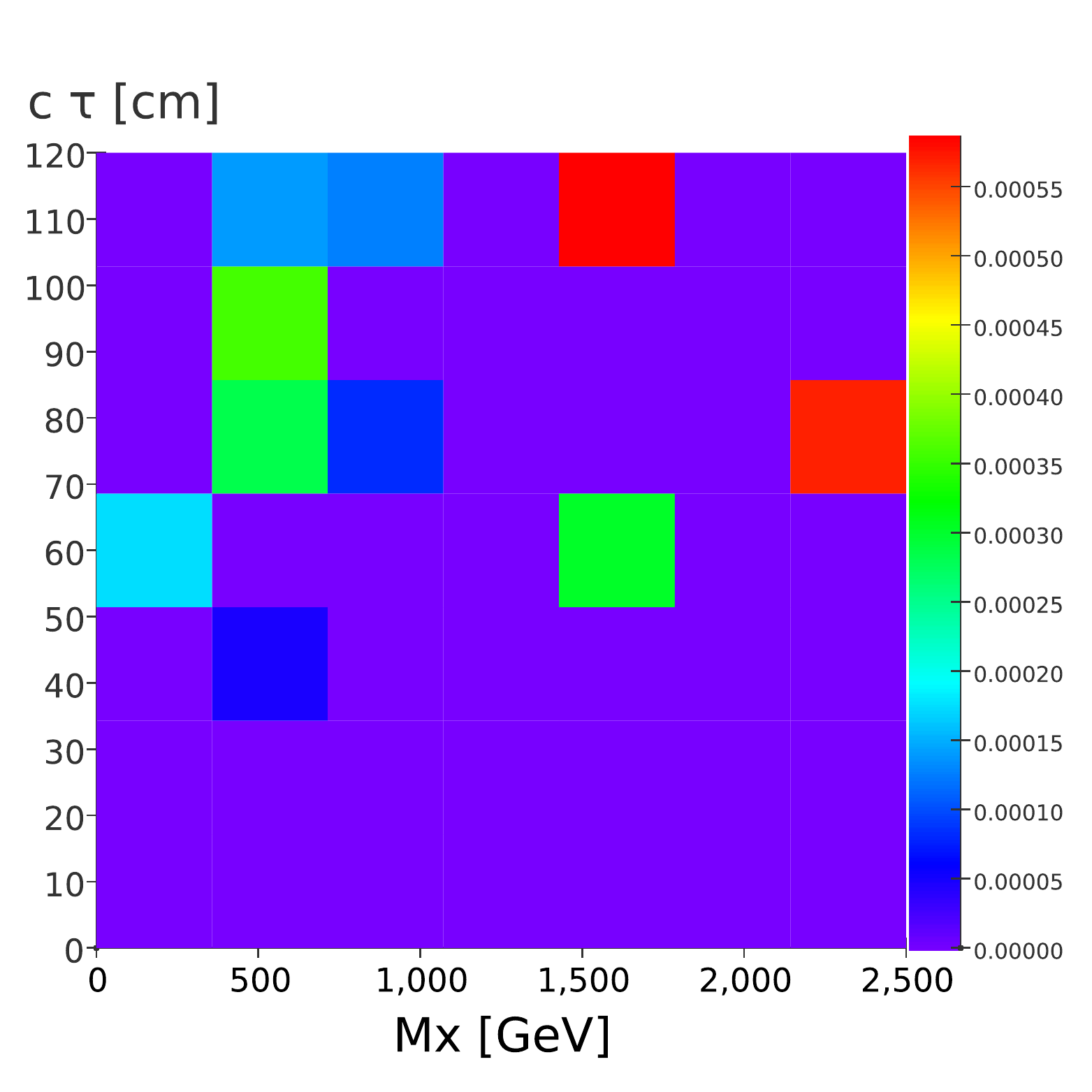}\hfill
   }

\end{center}
\caption{
The acceptance for emerging jets using timing layers with different timing resolutions as
a function of the mediator mass $M_X$ and the $c\tau$ of the dark pions with mass of 5~GeV.
The {\sc pythia}8 simulations were performed
for $pp$ collisions at $\sqrt{s}=13$~TeV. The maximum values for 
the color mapping
vary from  0.56 (a)  to  0.0006 (f).  
}
\label{fig:efficiency_med}
\end{figure}

The acceptance of the emerging jets was calculated as the fraction of  events that pass the 
Eq.~\ref{eqTOF} condition  with the parameters discussed earlier. 
 Figure~\ref{fig:efficiency_med} shows the acceptance as a function of the mediator mass $M_X$ and the decay distance
of the dark pions. This acceptance can be compared to the acceptance based on tracks~\cite{Sirunyan:2018njd}. 
The acceptance based on the TOF is significantly larger for low $M_X$ and large $L=c\tau$ of the dark pions.  
The acceptance is larger for the timing layers with a resolution better than 100~ps, as compared to the standard 1~ns resolution. 

We are interested in the acceptance  for dark pions  as a function of their mass
and lifetime assuming a fixed mass $M_X$ of the mediator. We  consider the HE-LHC environment 
with $pp$ collisions at a centre-of-mass energy of 27~TeV.    
The Monte Carlo generator settings for the signal model were similar
to those discussed in~\cite{Sirunyan:2018njd,prive}.
The mediator mass was set to 10~TeV, while the mass of the dark pion was varied in the range between 5~GeV and 1~TeV.
The dark pion decay length, $c\tau$,  was varied between 1~mm and 1000~mm, independent of its mass. Other parameters
were also appropriately modified to allow  sufficient phase space for the dark meson production.
The mass of the dark pion is assumed to be one half the mass of the dark quark. The mass of the dark
$\rho$ is four times the dark pion mass. The width of the
mediator particle is assumed to be small compared to the detector mass resolution.

\begin{figure}
\begin{center}
   \subfigure[10 ps] {
   \includegraphics[width=0.39\textwidth]{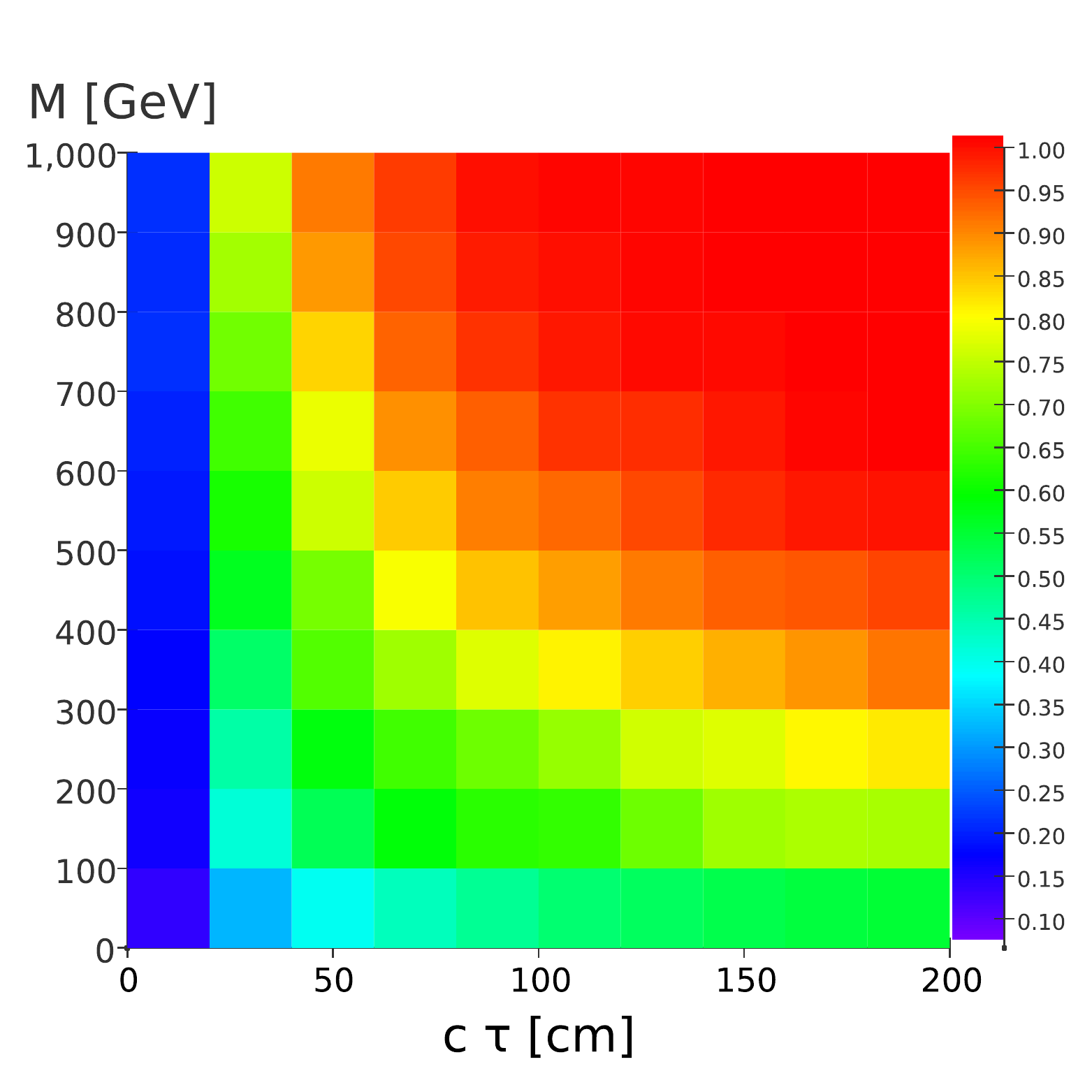}
   }
   \subfigure[20 ps] {
   \includegraphics[width=0.39\textwidth]{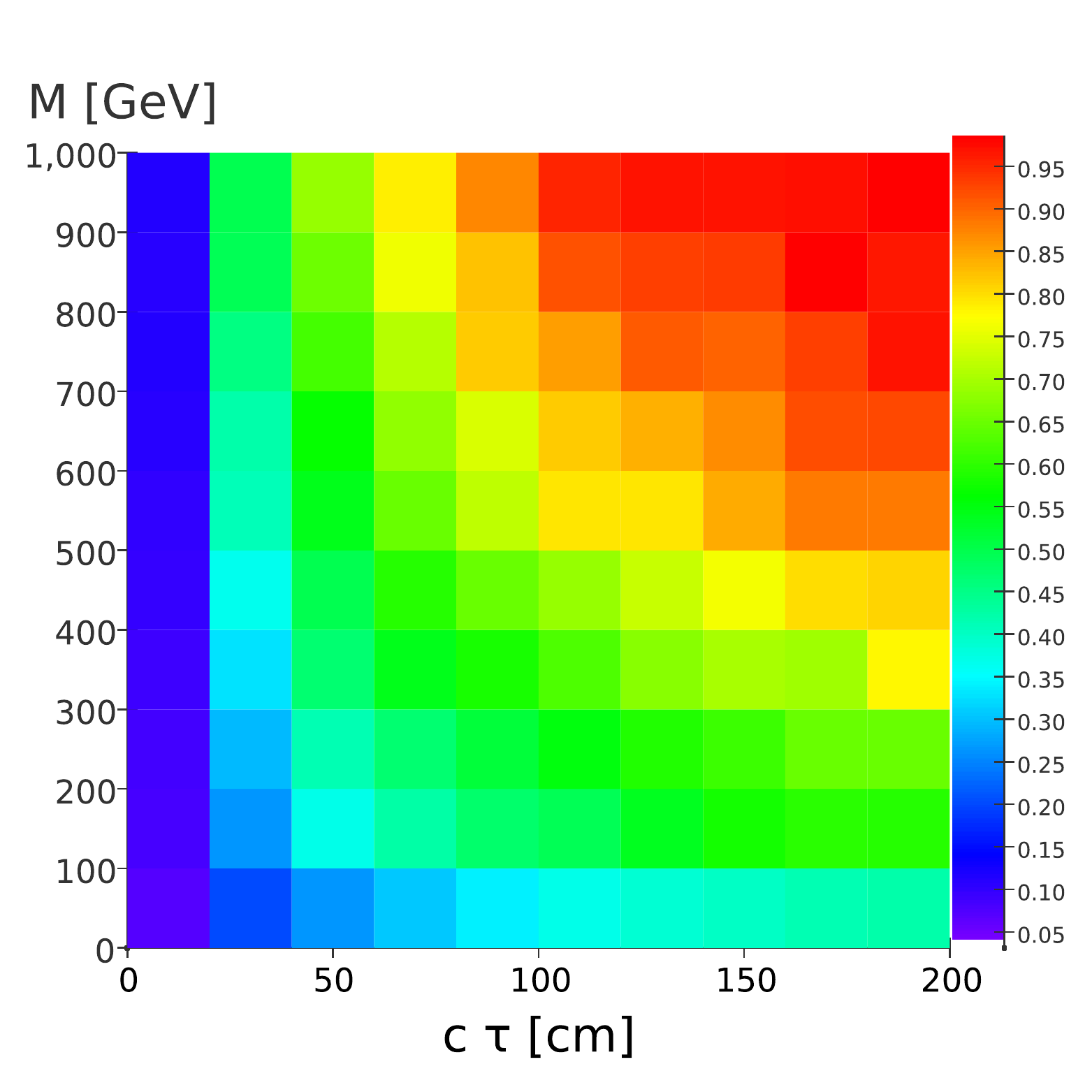}\hfill
   }

   \subfigure[30 ps] {
   \includegraphics[width=0.39\textwidth]{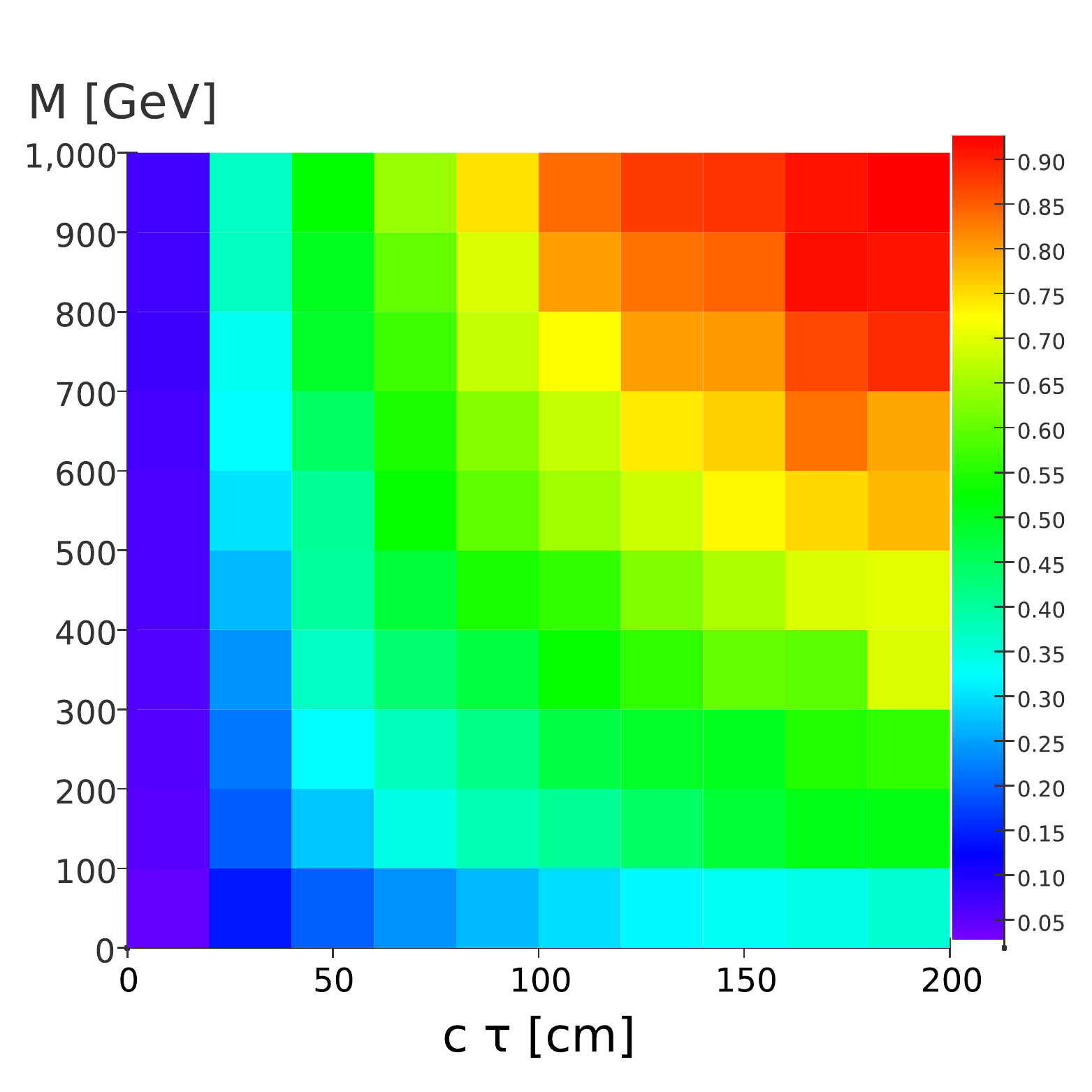}
   }
   \subfigure[100 ps] {
   \includegraphics[width=0.39\textwidth]{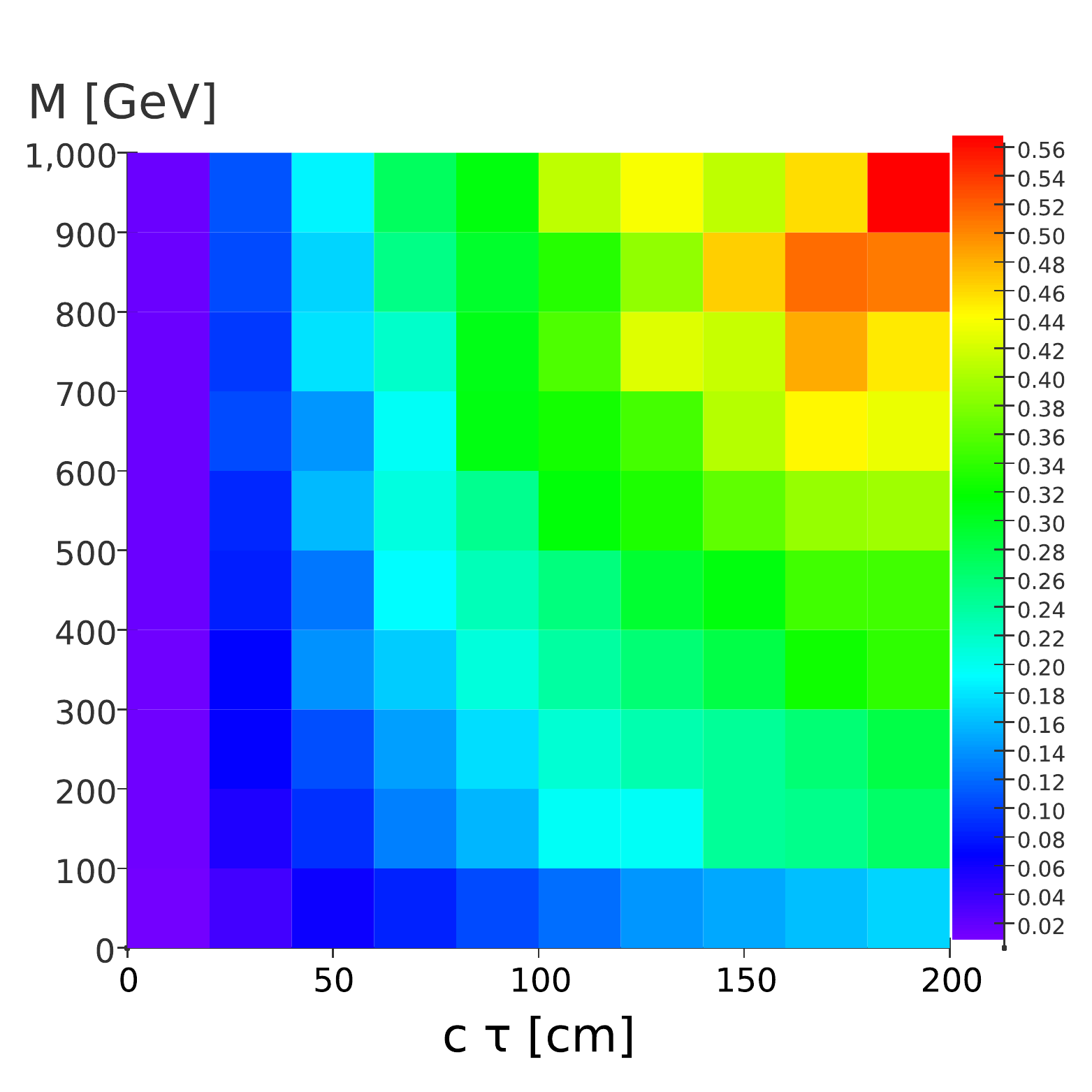}\hfill
   }

   \subfigure[500 ps] {
   \includegraphics[width=0.39\textwidth]{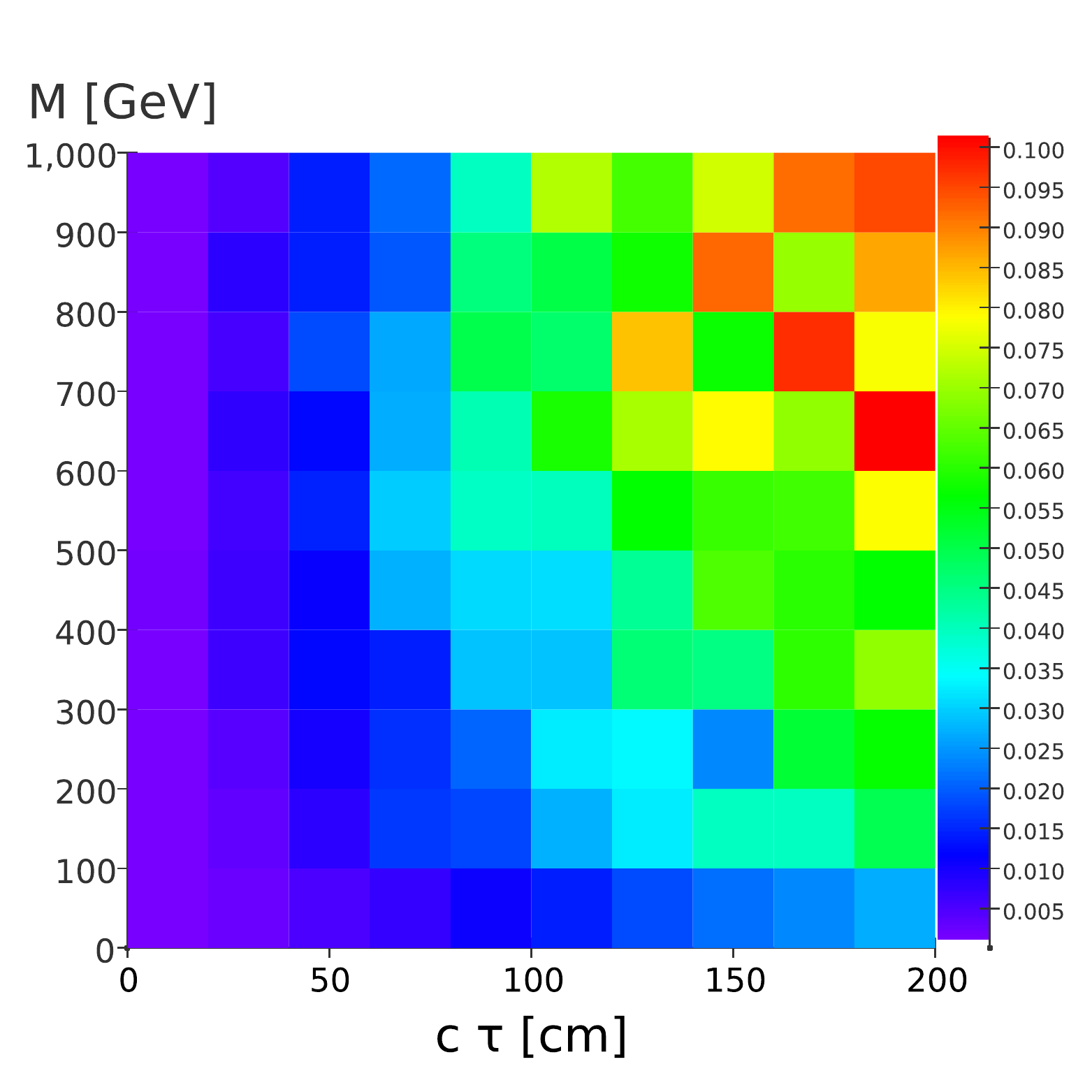}
   }
   \subfigure[1000 ps] {
   \includegraphics[width=0.39\textwidth]{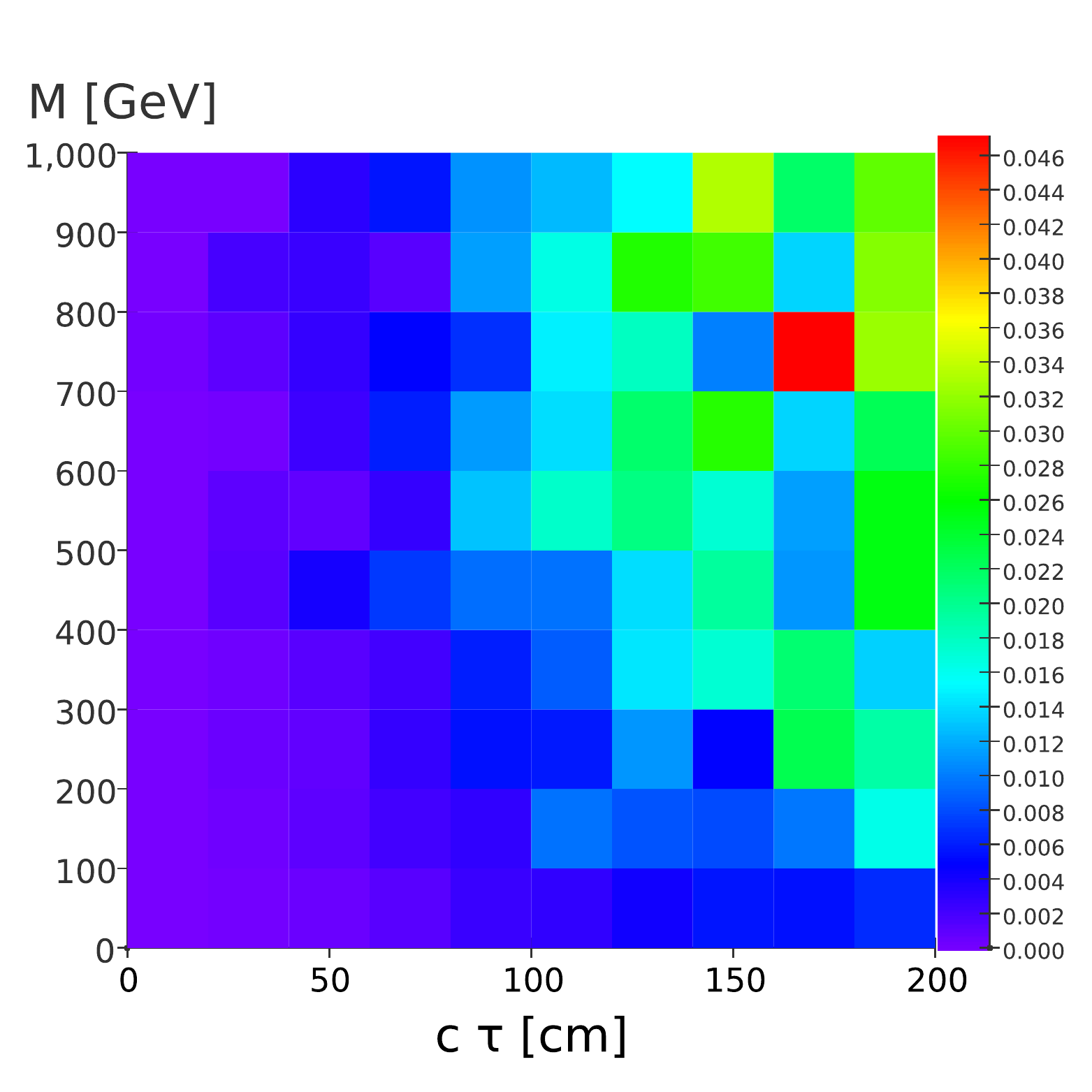}\hfill 
   }

\end{center}
\caption{
The acceptance for emerging jets using timing layers with different timing resolutions as
a function of the dark pion mass and $c\tau$. The mediator mass was fixed at $M_X = 10$~TeV. The {\sc pythia}8 simulations were performed 
for $pp$ collisions at $\sqrt{s}=27$~TeV. 
The maximum values for 
the color mapping 
vary from  1.0 (a)  to  0.046 (f).     
}
\label{fig:efficiency}
\end{figure}

As before, the acceptance for the emerging jets based on timing was calculated as the fraction of events that pass the 
Eq.~\ref{eqTOF} condition. Figure~\ref{fig:efficiency} shows the efficiency
as a function of $c\tau$ and the dark pion mass. It can be seen that a detector with the standard 1~ns resolution does
not have acceptance for the dark meson measurements. The acceptance is significantly larger when the timing layers 
 have a resolution better than 100~ps.
The acceptance is low for small $c\tau$ or small masses, which is the expected feature of the timing measurement.
The timing layers with 20~ps resolution have 100\% acceptance for large values of $c\tau$ and dark-meson masses.
The acceptance as a function of the particle velocity when using 20~ps and 1~ns resolution 
is shown in \ref{appendix}.

Note that these results are relatively general since they are formulated in terms of masses and decay lengths,  
i.e. independent of the position of the timing layers and other details relevant to 
the detector geometry.

\section{Summary}

This paper discusses the benefits of timing layers positioned in front of the hadronic calorimeters.
Using the full {\sc geant}4 simulation and a semi-analytic approach,
the figures of merit for the identification of single particles
using timing layers with resolutions of 10~ps -- 1~ns were calculated.
It was illustrated 
how such layers can be used for single-particle identification and
discrimination of heavy long-lived particles in the context of the dark QCD model.
It was shown that the timing layers lead to a significant benefit for reconstruction of heavy long-lived particles 
in the range of $c\tau$ and momentum where track-based measurements have low acceptance.

\section*{Acknowledgments}
The Monte Carlo events were generated using resources provided by the Open Science Grid, 
which is supported by the National Science Foundation award 1148698, and the U.S. 
Department of Energy's Office of Science.
We gratefully acknowledge the computing resources provided on a
high-performance computing cluster operated by the
Laboratory Computing Resource Center at Argonne National Laboratory.
The submitted manuscript has been created by UChicago Argonne, LLC,
Operator of Argonne National Laboratory (“Argonne”). Argonne, a U.S.
Department of Energy Office of Science laboratory, is operated under
Contract No. DE-AC02-06CH11357. The U.S. Government retains for itself,
and others acting on its behalf, a paid-up nonexclusive, irrevocable
worldwide license in said article to reproduce, prepare derivative works,
distribute copies to the public, and perform publicly and display
publicly, by or on behalf of the Government.
Argonne National Laboratory’s work was
funded by the U.S. Department of Energy, Office of High Energy Physics
under contract DE-AC02-06CH11357.

%%%%%%%%%%%%%%%%%%%%%% references %%%%%%%%%%%%%%%%%%%%%%%%%%%%%%
\section*{References}

\bibliographystyle{elsarticle-num}
\def\bibname{\Large\bf References}
\bibliography{biblio}

\clearpage

\appendix
%\renewcommand{\thesubsection}{\Alph{subsection}}
%\section*{Appendices}
%\addcontentsline{toc}{section}{Appendices}
\section{Acceptance vs $\beta$}
\label{appendix}

\begin{figure}
\begin{center}
   \subfigure[20 ps] {
   \includegraphics[width=0.41\textwidth]{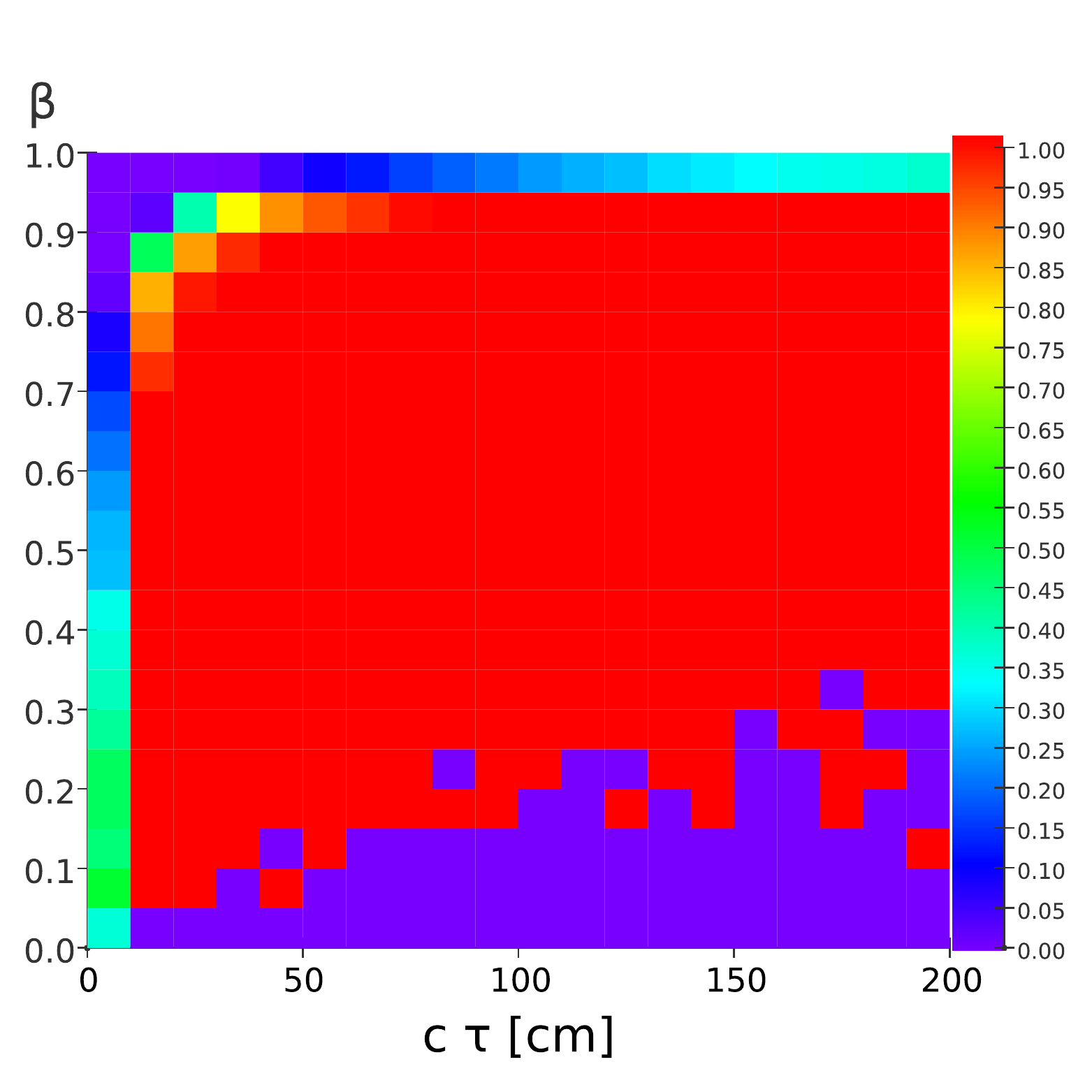}\hfill
   }
   \subfigure[1000 ps] {
   \includegraphics[width=0.41\textwidth]{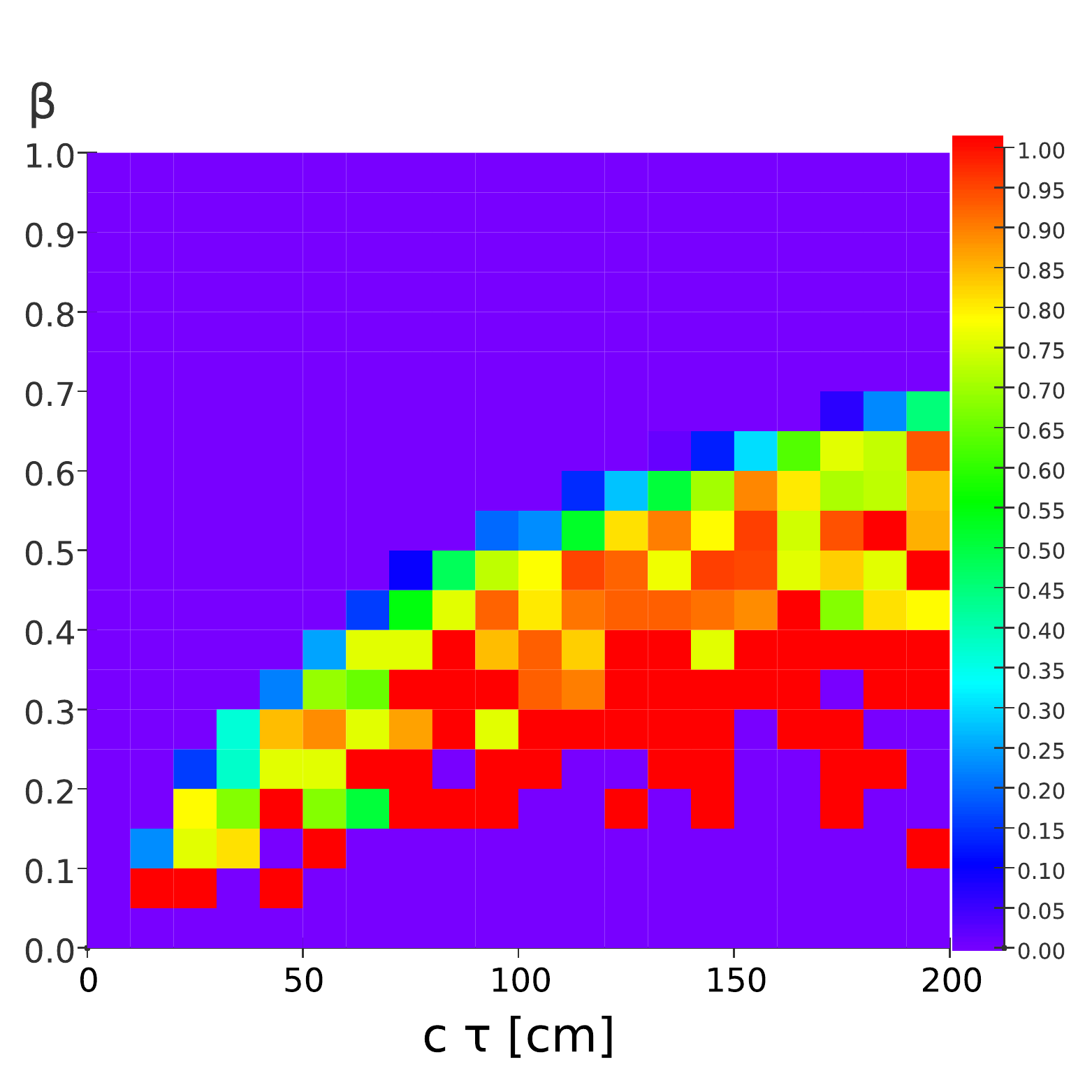}\hfill
   }

\end{center}
\caption{
Acceptance for the reconstruction of emerging jets using the timing layers with different timing resolutions.
The histogram shows the acceptance as a function of $c\tau$ and the  particle velocity $\beta$. 
}
\label{fig:efficiency_beta}
\end{figure}

Figure~\ref{fig:efficiency_beta} shows the reconstruction
acceptance as a function of $c\tau$ and the  particle velocity  $\beta=|p|/E$, for the two extreme
cases of the timing layers.
The calculations were performed using the Monte Carlo simulations for emerging jets (see the text).

\end{document}